\documentclass[aps,12pt,final,notitlepage,oneside,onecolumn,nobibnotes,nofootinbib,superscriptaddress,noshowpacs,centertags]{revtex4}

\newcommand{\const}{{\rm\, const}}

\usepackage{graphics}
\usepackage{bm}

\input epsf
\usepackage{amsmath,amssymb,epsfig,placeins,subfigure,wrapfig,indentfirst}

\begin{document}

\title{To the search for observational evidence of wormholes }

\author{Alexei Pozanenko}
\affiliation{Space Research Institute of the Russian Academy of
Sciences, 117997, Russia, Moscow,  Profsoyuzanaya, 84/32}

\author{Alexandr Shatskiy}
\affiliation{Astrospace Centre of the Lebedev Physical Institute
of the Russian Academy of Sciences, 117997, Russia, Moscow,
Profsoyuzanaya, 84/32}

\date{\today}

\begin{abstract}
We consider observational properties of gamma-ray bursts (GRB)
transmitted by hypothetical wormholes (WH). Such burst would be
observable as repeating source, analogous to Soft Gamma-Repeaters
(SGR). We show that the known sources of SGR cannot be WH
candidates. We also discuss observational properties of GRB which
might be a signature of WH.
\end{abstract}


\maketitle

\section{Introduction}
\label{s1}

Papers by Kardashev, Novikov, and Shatskiy~\cite{Kardashev06,
Kardashev07} as well as by Shatskiy~\cite{Shatskiy07}  deal with a
hypothesis that some astrophysical objects (for example, some
active galactic nuclei or quasars) are supposedly entrances to
wormholes (WH). A wormhole in general relativity is a topological
tunnel connecting two distant parts of the universe space (see,
for instance,~\cite{Einstein37, Misner57, Morris1988, Visser95,
Ellis73, Frolov98}) or even two parts of different universes in
the Multiverse model~\cite{Carr07}).

Indeed, if radiation from another universe comes through to our
Universe a distant observer will perceive a WH throat as a point
source. Both the invariable radiation of the sources of the other
universe (stars and galaxies) and the radiation of transient
sources will be transmitted to our universe. In the latter the
brightest known sources are Gamma-Ray Bursts (GRB). Their
brightness in the gamma-rays  may exceed for a short time interval
that of the entire sky. If GRB phenomena are presumably present in
the other universe (the fact that there are GRBs located in
distant parts of our universe is undoubted -- having been detected
up to z $\sim$8.2 \cite{z8.2}, they are the farthest objects in
the Universe which is directly observable now), an observer in our
universe will register repeated aperiodic gamma-ray flashes coming
from a single point spatially coinciding with a WH.

Such sources with similar properties of repeating radiation do
exist. These are Soft Gamma-ray Repeaters (SGR). Nowadays most
known SGRs can be reliably associated with magnetars in the Galaxy
(e.g. \cite{magnitar}). However, observational signatures from
GRBs of the other universe can still be reminiscent of some
properties of SGRs.

In this paper we study possible observational signatures from GRBs
of the other universe transmitted by hypothetical wormholes.

\section{Derivation of differential power spectrum (continuous approach)}
\label{s3-0}

One of few quantitative features we can estimate for GRB
transmitted to our universe is the cumulative distribution ${\log
N(>s)-\log s}$, i.e. the number of sources $N$ with flux exceeding
$s$. The distribution can be referred to as an integral power
spectrum (IPS). Hence, a differential power spectrum $N'(s)$ is
related to the integral one through the formula ${N(s)=\int
N'(s)ds}$.

In Euclidian 3-dimensional space the power law index of the
differential power spectrum (DPS) of uniformly distributed
identical sources is $n=-5/2$:
\begin{eqnarray}
N'(s)\propto s^{-5/2} \label{2-1}\end{eqnarray} Bright GRBs yield
the integral power spectrum index to be $\sim -3/2$
(e.g.~\cite{BATSE1})

The number $N'(s)$ is in direct proportion with probability
density $P$ for $s$ photons to fall onto the WH throat,
${N'(s)\propto P(s)}$.

% The spectra DPS and IPS in the other universe can be supposed to be the same.

A more straightforward and illustrative way of calculating the
spectra is the one that has to do with averaging over all the
sources in unit volume and integrating over all the photons
emitted by them per time unit. Indeed, in the gamma range where
GRBs were first discovered and are currently being detected (30 --
1000~keV) the number of photons emitted by a single GRB is
sufficiently large.

However, in the GeV -- TeV range where the number of photons
emitted by the sources is merely a few we must take advantage of
the discrete approach set out in Section~\ref{s3}. In the limit of
high intensity of source flux the two techniques lead to the same
result.

Consider a spherically symmetric case with a wormhole connecting
our universe and the other. Let $4\pi R^2$ be the area of the
sphere, which an observer rests on, and $4\pi r^2$ be the area of
the sphere, which holds a source (of the other universe). Let
$\theta$ be an angle in the spherical coordinates of the other
universe, at which the source is positioned\footnote{The angle
${\theta =0}$ corresponds, by definition, to a light ray moving
through the center of the throat (the ray with zero impact
parameter, see~\cite{Shatskiy2009}).}.

Let ${I}$ be the intensity of the source in the other universe.
Consequently, the number $s$ of the photons detected per unit time
by the observer in our universe with detector of the surface $S_1$
is given by
\begin{eqnarray}
s= I\cdot\frac{S_0}{4\pi r^{2}}\cdot f(\theta)\cdot\kappa
\cdot\frac{S_1}{4\pi R^{2}} \label{s-1}
\end{eqnarray}
Here ${S_0\equiv 4\pi r_0^2}$ is the effective area of the WH
throat surface and ${f(\theta)}$ is the deflection function that
defines how the apparent brightness of the source detected by the
observer (in our universe) changes with angle $\theta$ (and/or
with impact parameter in the throat) while the factor ${\kappa
=const}$ describes photon losses on passing through the throat.
Moreover, since we adopt that the factor $\kappa$ accounts for the
losses, the integral over the entire solid angle ${d\Omega
=2\pi\sin\theta\, d\theta}$ with ${f(\theta)}$ being the integrand
yields ${4\pi}$.

Assume that the sources are identical ${(I=const)}$ and uniformly
distributed. The mean density\footnote{The averaging is performed
over the volume that contains many sources and still much smaller
than $r^3$.} of the sources in the other universe is ${\rho
=const}$. Let ${\omega =const}$ be the average rate of the events
per unit volume and time and $T$ the duration of observation
\footnote{The averaging is performed over the time interval $\tau$
which satisfies the inequality ${2\pi /\omega <<\tau <<T}$.}.
According to eq. (\ref{s-1}), the condition ${s=const}$ defines a
surface around the throat, on which holds the relation:
\begin{eqnarray}
r_{s}^2(\theta) = \frac{I S_0 \kappa S_1}{16\pi^2 R^2}
\cdot\frac{f(\theta)}{s} \label{s-2}
\end{eqnarray}
Then the number $N$ of  detected events with photon number
exceeding ${s}$ (that came from the entire solid angle of the
other universe within the time $T$) is given by the integral with
respect to volume from the throat to the surface:
\begin{eqnarray}
N(>s)=\omega T\rho\, \int\limits_0^{\pi}2\pi\sin\theta\, d\theta\,
\int\limits_{r_0}^{r_s(\theta)}\, r^2\, dr = \frac{2\pi\omega
T\rho}{3}\, \int\limits_0^{\pi}\, \left[
r_s(\theta)^3-r_0^3\right]\,\sin\theta\, d\theta\, , \label{N-1}
\end{eqnarray}
where the function ${r_s(\theta)}$ is defined by eq.~(\ref{s-2}).
From the last formula we obtain:
\begin{eqnarray}
N(>s)=\frac{2\pi\omega T\rho}{3} \left(\frac{I S_0 \kappa
S_1}{16\pi^2 s R^2}\right)^{3/2}\, \int\limits_{-1}^{1}
f^{3/2}(t)\, dt -\frac{4\pi\omega T\rho r_0^3}{3} =const\cdot
s^{-3/2}-const\, , \label{N-2}
\end{eqnarray}
where we introduced the substitution ${t\equiv\cos\theta}$.

And we finally obtain the DPS:
\begin{eqnarray}
\frac{dN}{ds}=const\cdot s^{-5/2} \label{N-3}
\end{eqnarray}
This coincides with expression (\ref{2-1}), which could be
speculated to be true due to the spherical symmetry of the case.

\section{Derivation of differential power spectrum (discrete approach)}
\label{s3}

After passing through the WH throat the GRB photons will tend to
be redistributed inside a solid angle in accordance with equations
of papers \cite{Shatskiy2004, Shatskiy2009} rather than uniformly.

\begin{figure}
\includegraphics[width=0.9\textwidth]{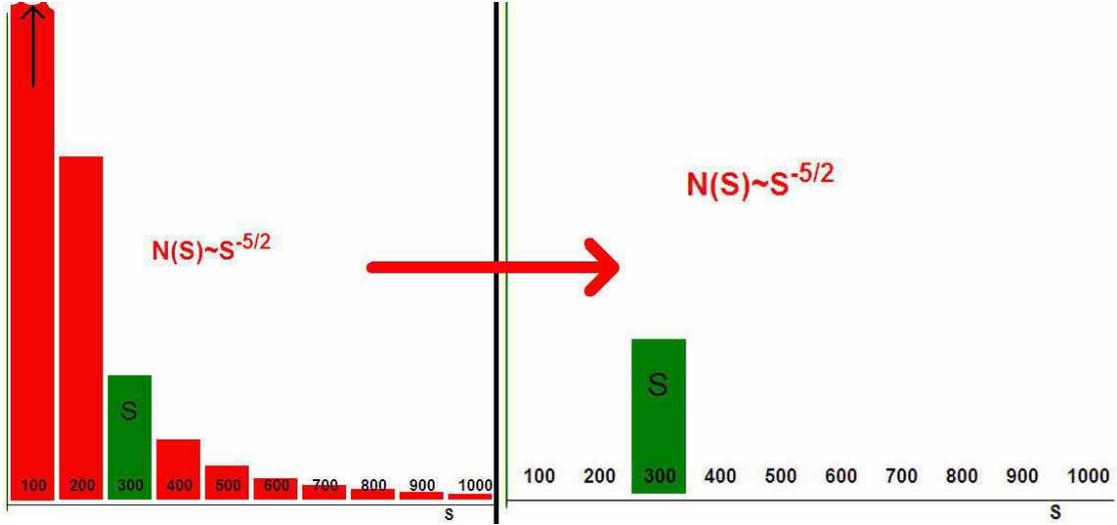}
\caption{{Using a portion of the full DPS (green column) to derive
how the WH throat distorts that portion.} \label{R2-5-1}}
\end{figure}

To begin with, we calculate the probability distribution $P^s_k$
which defines the probability to detect $k$ photons (that reached
the observer's telescope) of the maximal $s$ photons that had been
emitted by a GRB and passed through the WH throat. To do so, we
apply the following assumptions (see Fig.~\ref{R2-5-1} on
p.~\pageref{R2-5-1}):

1. Let the other universe contain ${N_s}$ stars of equal
luminosity, ${N_s>>1}$.

2. The throat transmits $s$ photons per unit time from every
single star.

3. Let the stars be uniformly distributed over the celestial
sphere of the other universe.

The observer in our universe looking at the stars in the other
universe through the WH throat sees their uneven distribution in
the throat. This is because the WH throat deflects and distorts
the light of the stars. It is obvious that the distortion will be
spherically symmetric and centered around the center of the WH
throat.

Now, let the observer look only at the portion of the stars within
a thin ring co-axial with the throat, $h$ and $dh$ being its
radius and thickness. With that the observer examines the solid
angle ${d\Omega (h)}$ of the other sky, ${d\Omega = 2\pi
|\sin\theta |\, d\theta}$. Since the full solid angle equals to
${4\pi}$, in the ring the observer counts ${dN_s=N_s\, d\Omega
/(4\pi)}$ stars\footnote{Since the light deflection angle $\theta$
can be more than $\pi$, the full solid angle turns out to be more
than ${4\pi}$. This substitution, however, reduces to another
constant (instead of ${4\pi}$) and does not affect the final
result.}. That is, the apparent density of the stars (the number
per unit area of the ring) appears to be
${J=\frac{dN_s}{dS_h}=\frac{N_s}{4\pi}\cdot\frac{d\Omega}{dS_h}}$,
where ${dS_h\equiv2\pi h\, dh}$. Since in our model all $N_s$
stars simulate all possible positions of the identical GRBs (with
the number $s$ of photons that passed through the WH), the value
of the apparent star density $J$ is to be in direct proportion
with the sought probability, ${J\propto P^s_k}$.

The distortion of light rays that passed through the WH throat is
caused not only by redistribution of the apparent star density,
but also by the fact that their apparent brightness undergoes a
change, viz. the brightness changes with increase of the impact
parameter $h$. This is because increasing the radius $h$ of the
ring, which transmits the starlight, changes a solid-angle
element, which the light scatters into. The respective apparent
stellar brightness proportional to the number $k$ of detected
photons is in direct proportion with ${\frac{dS_h}{d\Omega}}$ and,
thus, ${k\propto 1/J}$. This yields the sought probability:
\begin{eqnarray}
P^s_k=\frac{C(s)}{k} \label{3-4}
\end{eqnarray}
Here the multiplicative factors $C(s)$ of the probability are
defined by the stellar brightness (or the magnitude of a GRB),
i.e. by the number $s$ of GRB photons that passed through the WH.
To obtain the factors ${C(s)}$ we use the condition of probability
normalization:
\begin{eqnarray}
\sum\limits_{k=1}^s P_k^s=C(s)\cdot\sum\limits_{k=1}^s
\frac{1}{k}=1 \quad\Rightarrow\quad C(s)=\left(\sum\limits_{k=1}^s
\frac{1}{k}\right)^{-1} \label{3-4-1}\end{eqnarray}

As ${N_s\to\infty}$, the apparent mean brightness of a region
within the WH throat is independent of the impact parameter and
the wormhole looks like a homogeneous spot in every wavelength
range, regardless of a WH model\footnote{As ${N_s\to\infty}$,
separate stars become unseen -- they blur because of the finite
resolution of an observing device. This leads to the averaged
overall brightness of the stars being independent of $h$, i.e. the
brightness distribution is uniform. On the contrary, if the value
of $N_s$ is kept down one can see single stars, i.e. the
distribution becomes uneven. Remarkably, this result is universal:
it holds for every WH model, provided the total mass of the WH is
non-negative.}.

\begin{figure}
\includegraphics[width=15cm]{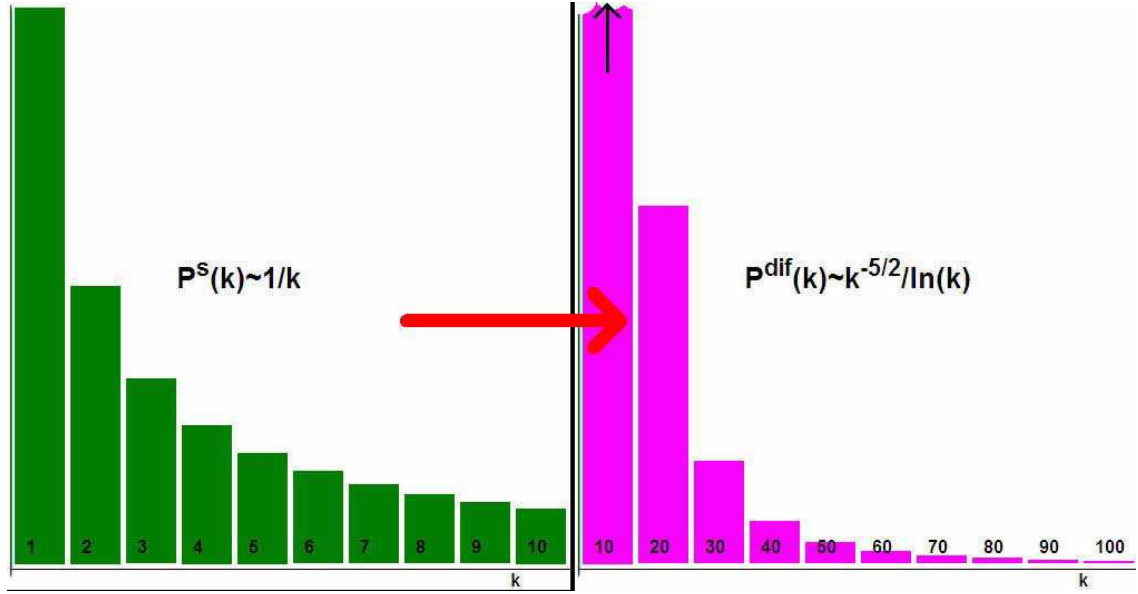}
\caption{{Transition from $P_k^s$ to $P_k^{dif}$ as a sum of
probabilities of every possible $s$ with weights proportional to
$s^{-5/2}$.}} \label{R2-5-2}
\end{figure}

According to eq. (\ref{2-1}), the set of identical GRBs $j(s)$
(with equal numbers $s$ of photons that passed through the WH) is
distributed as follows:
\begin{eqnarray}
j(s)=\const\cdot s^{-5/2} \label{3-4-3}
\end{eqnarray}
Therefore, the probability to detect $k$ photons from all $j(s)$
parcels with $s$ photons each is newly re-normalized sum of
probabilities:
\begin{eqnarray}
P_k^{js}=\const\cdot\sum\limits_{i=1}^{j(s)} P_k^s =
\const\cdot\frac{s^{-5/2}\cdot C(s)}{k} \label{3-4-2}
\end{eqnarray}
Thus, the index of the DPS from GRBs, which are equally bright
near the WH throat (with the number ${j(s)\cdot s}$ of photons
that passed through the WH and the number $k$ of those of them
that reached the detector), is equal to ${-1}$.

Therefore, the sought-for total probability $P_k^{dif}$ to detect
$k$ photons from every possible GRB (with different $s$) is a
re-normalized sum of probabilities ${P_k^{js}}$ for every possible
value of ${s\ge k}$:
\begin{eqnarray}
P_k^{dif}=\const\cdot\sum\limits_{s=k}^\infty P_k^{js} =
\frac{\const}{k}\cdot\sum\limits_{s=k}^\infty
\cdot\frac{C(s)}{s^{5/2}} \label{3-5}
\end{eqnarray}
In other words, the total probability $P_k^{dif}$ is the sum of
probabilities $P_k^s$ from every possible GRB with weights, which
are in direct proportion with distribution (\ref{2-1}) (see
Fig.~\ref{R2-5-2} on p. ~\pageref{R2-5-2}).

\begin{figure}
\includegraphics[width=15cm]{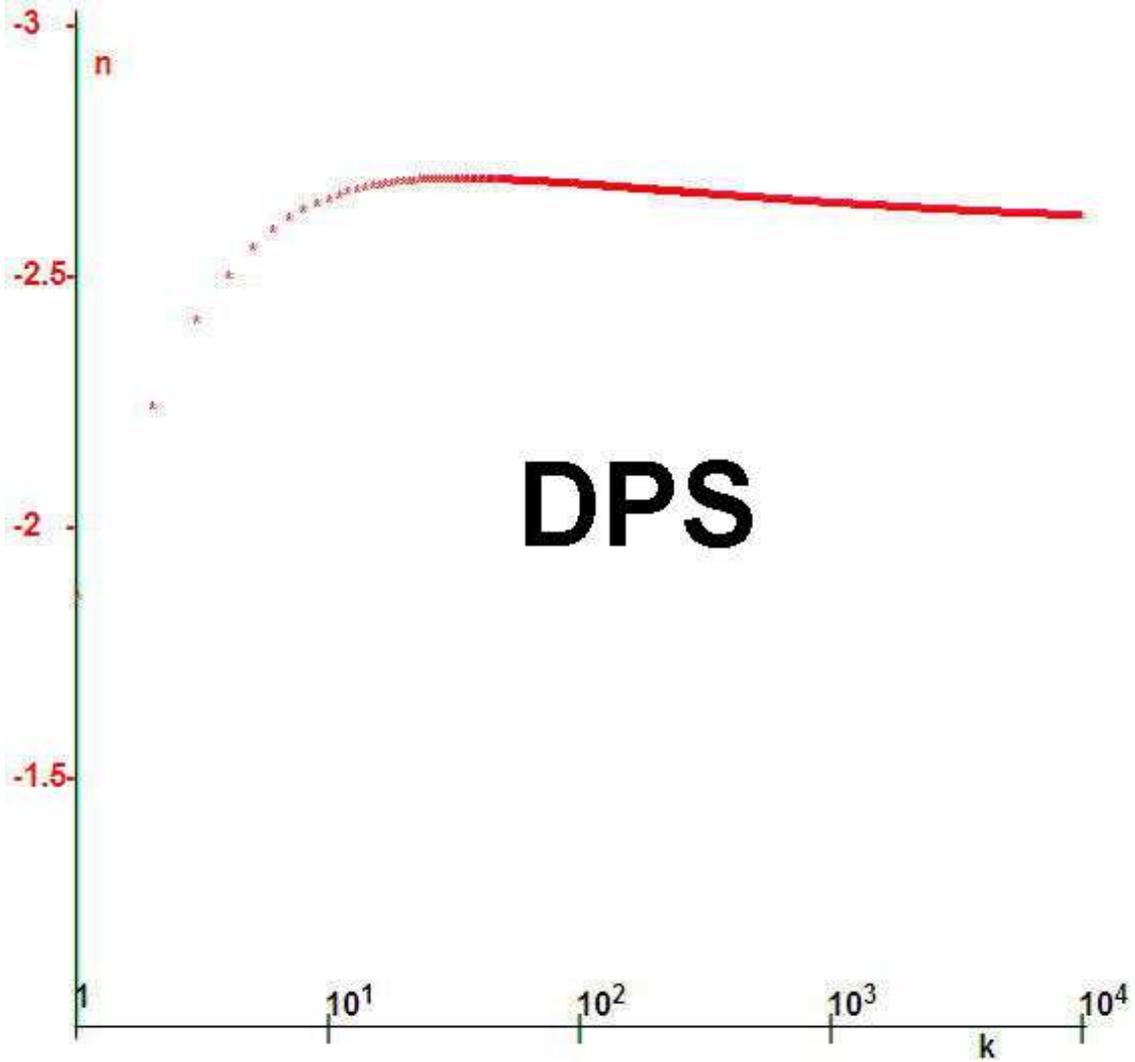}
\caption{{Numerical values of the discrete function ${n(k)}$
giving the DPS  for a WH in the range ${1\le k\le 10^4}$. The
deviation of the power law index from $-5/2$ has to do with edge
effects near the WH throat and the function $N'(s)$ being singular
at zero, see~(\ref{N-3}).}} \label{R2-5-3}
\end{figure}

The DPS index ${n(k)}$ is given by the expression:
\begin{eqnarray}
n(k)\equiv\frac{d\ln P^{dif}_k}{d\ln k}=
\frac{k}{P^{dif}_k}\cdot\frac{dP^{dif}_k}{dk}
\label{3-9}\end{eqnarray} Using expressions  (\ref{3-4-1}),
(\ref{3-5}) and (\ref{3-9}), we can write down the exact
expression for the DPS index ${n(k)}$:
\begin{eqnarray}
-n(k)=1+\frac{C(k)}{k^{3/2}\left[C_1-\sum\limits_{s=1}^{k-1}
s^{-5/2}\cdot C(s) \right]}\, , \makebox{ where
}C_1\equiv\sum\limits_{s=1}^{\infty} \frac{C(s)}{s^{5/2}} \approx
1.1933853736824342 \label{3-11}\end{eqnarray} Figure~\ref{R2-5-3}
shows the discrete function $n(k)$ plotted exactly for small
values of $k$, for large values $k$: ${n\approx -5/2}$ -
see~(\ref{N-3}).

\begin{figure}
\includegraphics[width=15cm]{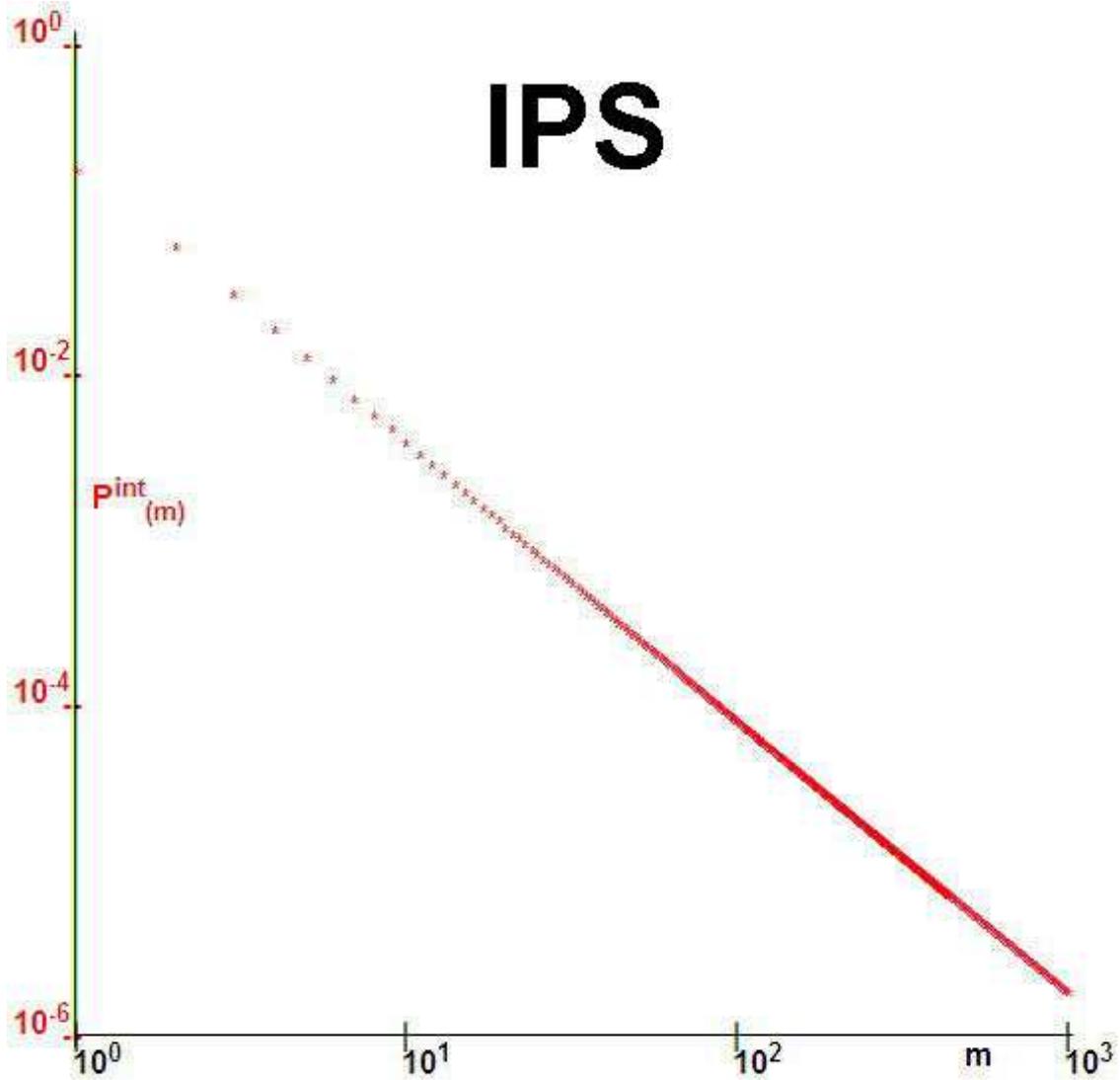}
\caption{{Numerical values of the discrete IPS function of a WH in
the range ${1\le m\le 10^3}$. The IPS spectral index ${n\approx
-1.67}$.}} \label{R2-5-4}
\end{figure}

%\subsection{Calculation of differential power spectrum (discrete approach)}
%\label{s4}

In order to obtain the DPS, we merely require to sum up expression
(\ref{3-5}) with respect to $k$ from $m$ to $\infty$.

Summing up eq. (\ref{3-5}) yields:

\begin{eqnarray}
P_m^{int}=\const \sum\limits_{k=m}^\infty \left(\frac{1}{k}
\sum\limits_{s=k}^\infty \frac{C(s)}{s^{5/2}} \right)= \const
\sum\limits_{k=m+1}^\infty \left(\frac{1}{k}
\sum\limits_{s=k}^\infty \frac{C(s)}{s^{5/2}} \right)+
\frac{\const}{m}\sum\limits_{s=m}^\infty \frac{C(s)}{s^{5/2}}=
\nonumber\\
=P_{m+1}^{int}+P^{dif}_m \label{4-2}
\end{eqnarray}
or
\begin{eqnarray}
P_{m+1}^{int}=P_{m}^{int}-\frac{\const}{m}\left[ C_1-
\sum\limits_{s=1}^{m-1} \frac{C(s)}{s^{5/2}}\right] \label{4-3}
\end{eqnarray}
Using this recurrent relation it is possible to plot the exact IPS
(see Fig.~\ref{R2-5-4}).

\section{Can we observe GRB transmitted by wormhole?}
\label{s5}

\begin{figure}
\includegraphics[width=15cm]{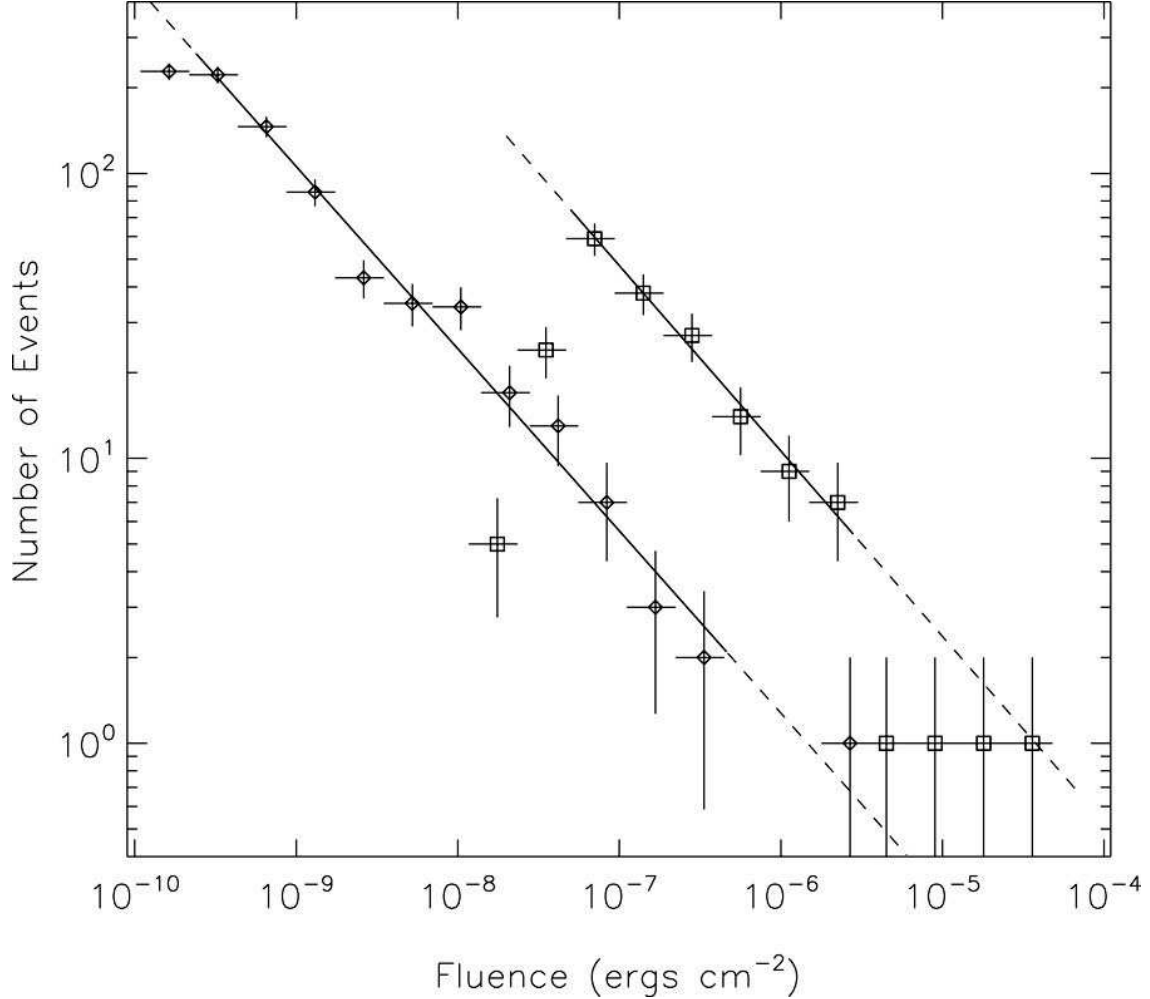}
\caption{{Differential spectrum  (${\log N(s)-\log s}$) of
SGR~1900+14. The spectral index is ${n\approx -1.66}$.} Squares
denote BATSE experiment, diamonds -- RXTE  \cite{Gogus1999}.}
\label{R2-5-5}
\end{figure}

Suppose that the properties of GRBs in the other universe are
analogous to those of GRBs in our universe. The brightest GRBs
detected so far provide the flux of $\sim {10^{-3}\, erg/(cm^2\,
s)}$ in the range ${30\div 300\, keV}$. On the other hand, the
sensitivity threshold of contemporary gamma-ray telescopes is
${\sim 10^{-8}\, erg/(cm^2\, s)}$ while significantly more
sensitive X-ray telescopes can detect fluxes up to ${\sim
10^{-12}\, erg/(cm^2\, s)}$. Let us estimate the distance $R$ of
the brightest GRB in the other universe, which, on having been
transmitted by a WH to our universe, would still be detected.

The flux $s_2$ received by an observer on the Earth located at the
distance $R$ from the WH is given by ${s_2 = s_1 \pi R_c^2/(4\pi
R^2)}$, where $R_c$ is the capture radius of the WH and $s_1$ the
GRB flux coming onto unit area of the WH throat in the other
universe. We also assume that photons losses on passing through
the WH vanish. Then, if the capture radius is ${R_c=4GM/c^2}$ (for
the case of the magnetic WH\footnote{In the general case we have:
${R_c=min[r/\sqrt{g_{tt}(r)}]}$ --- see.~\cite{Shatskiy2009}.}) we
obtain ${R = 0.5\sqrt{s_1/s_2} R_c = \sqrt{s_1/s_2} M/M_\odot
10^{-13}\, pc}$, where $M_\odot$ is the solar mass and $M$ the WH
mass. Even with extreme conditions being taken into account (the
highest GRB flux, the most sensitive up-to-date detectors (see
above) and a supermassive WH, ${M=10^{9}M_\odot}$), we obtain the
distance of mere 3 parsecs. It is evident that within a few
parsecs from the Earth there are no bodies that massive.

Yet, if the mass of the WH throat in our universe is much lower
than that of the WH throat in the other universe, such an object
may well be near the Earth (for example, a WH throat could
 be a single black hole of stellar mass) and
observations of GRBs of the other universe (alternatively, a
distant part of our own universe) could be possible at the present
time.

Besides, lensing may play a significant role. This implies that
the deflection function $f(\theta)$ has maxima in certain
directions (angles $\theta$), in which the apparent brightness of
a transmitted source is amplified \cite{Shatskiy2009}. If this is
the case, observations of GRBs of the other universe can finally
be carried out at essentially larger distances between the WH and
the observer. In this case, however, no repeating flashes will be
detected, since the lensing is realized only for a certain
geometry of the system GRB source -- WH -- observer.

Note the observational properties   of a WH in our universe. The
WH throat has to be positioned within a few parsecs from the
Earth, manifest itself in the gamma-ray range as a point source of
repeating gamma-ray bursts with integral spectrum index ${\log
N-\log s ~\sim -3/2}$, whereas hardness of the repeating gamma-ray
bursts and its duration depend on the mass ratio of the WH throats
in the other universe and our own realm.

Let us find out whether known soft gamma repeaters suit to be WH
candidates. Discovered in 1979~\cite{mazets1979}, sources of SGR
sporadically emit short (${\sim 0.1\, s}$) bursts of soft
gamma-radiation with the OTTB spectrum $\sim E^{-1}exp(-E/kT)$,
where $kT \sim 20-30$ keV (e.g.~\cite{SGRreview}), and, very
rarely, giant bursts. The best-known one of the latter was
detected from SGR1806-20 in 2004 on the 27th of
December~\cite{hurley_gaint}. Sources of SGR  are currently
believed to be magnetars, i.e. single neutron stars with strong
magnetic field (${\sim 10^{13}\div 10^{15}\, G}$).

Intensity distribution of SGR outbursts was studied, for example,
in papers by Gogus et al. \cite{Gogus1999, Gogus2000} (see
Fig.~\ref{R2-5-5}) where the authors showed that the differential
distribution ${\log N(s)-\log s}$ for SGR1900+14 is well fit by
the power law ${1.66 \pm 0.13}$ while SGR1806-20 requires the
power law index ${1.43 \pm 0.06,\, 1.76\pm 0.17}$ and ${1.67 \pm
0.15}$ corresponding to the experiments RXTE/PCA, BATSE and ICE,
respectively (see also \cite{gotz1806-20}). In both cases of
SGR1900+14 and SGR1806-20 the indices within the error bars are
inconsistent with $-5/2$ expected for a GRB transmitted through
WH.

Now, let us compare statistics of bursts durations and intervals
between them (i.e. waiting time between successive bursts) for
SGRs and GRBs. For SGR 1900+14 and SGR 1806-20 the statistics
yields typical interval between bursts of 100 s \cite{Gogus1999}
and 50 s \cite{Gogus2000}, respectively, and burst durations of
$\sim 0.16$ and 0.09 s \cite{Gogus2001}.

\begin{figure}
\includegraphics[width=15cm]{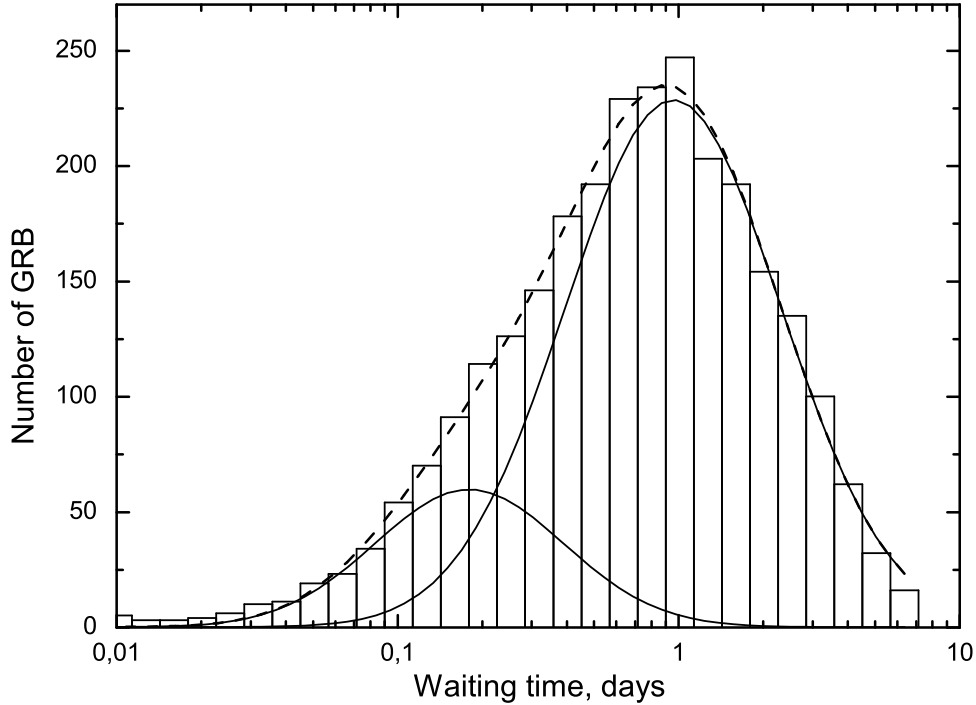}
\caption{Waiting time distribution for GRB in the BATSE experiment
(histogram),  two log-normal curve fitting the distribution
(solid), and sum of the curve (dashed). All
 events from the current BATSE GRB catalogue~\cite{BATSE_current_catalog} are used.}
\label{BATSE_delay}
\end{figure}

The characteristic (most probable) duration $T_{90}$ in the group
of long bursts ($T_{90} > 2$ s) is $\sim 30$ s \cite{kouveliotou}.
Figure~\ref{BATSE_delay} shows the distribution of waiting time
between bursts in the BATSE experiment, which along with the
analogous distribution for SGRs can be approximated by a
log-normal distribution with the most probable value of waiting
time being 0.95 days. The ratio of waiting time to event duration
is 550 -- 630 for SGRs and 3000 for GRBs. When a GRB is
transmitted through a WH the ratio of typical duration to waiting
time is not to be changed.

Indeed, the waiting time distributions (see
Fig.~\ref{BATSE_delay}) are adequately fit by two log-normal laws
(${\chi^2/d.o.f. = 1.5}$).   Since the field of view of the BATSE
experiment is somewhat less than ${4\pi}$, the observed
distribution is slightly shifted towards longer times than we
would expect from a distribution from the entire sphere. A
short-time tail could be caused by including to the catalogue
unidentified SGR events as well as by multiple
  triggering by the same burst, e.g. the precursor and the
burst itself or ultra-long GRBs detected as individual events.

As soon as duration and hardness are concerned, note also a
contradiction between SGR and GRB spectra. As it is known, most of
the observed GRBs are long duration bursts ($T_{90}>2$~s in the
BATSE experiment) and maxima in their spectral energy distribution
tend to be 300 keV~\cite{preece_catalog}. On the other hand, SGRs
have short durations and very soft spectra. The contradiction
could be partially resolved if, for some reason, we observe
through WHs only short duration gamma-ray bursts ($T_{90}<2$~s,
and whose durations start from a few milliseconds) forming a
separate class of events. But in this case the hardness difference
in the spectra turns out to be even sharper than when we compare
long GRBs and SGRs.

For some SGRs distance estimates are available, e.g. for
SGR1806-20 the distance to the source is estimated to be from 6
~\cite{cameron_1806} to 15 kpc~\cite{Corbel, McClure} and for
SGR1900+14 -- 7 kpc~\cite{Vasisht}. However, it is not the
distance estimated rather vaguely that prevents us from
associating known SGRs with wormholes. Permanent pulsating
radiation with periods ranging from 2 to 8 s demonstrated by most
SGRs, which can be easily explained in the magnetar model by
rotation of the neutron star, is unlikely to fit the current WH
models. One can firmly suggest that known SGRs can not be
considered as candidates of wormholes.

The search for recurrent events from classical GRBs was in vain.
The most probable candidate could be suspected in 4 successive
GRBs with overlapping localization error boxes, which were
detected by BATSE in October 1996~\cite{graziani} (BATSE triggers
5646, 5647, 5648, 5649). However, estimates of probability for
these to be actual recurrent bursts are highly dependent of the
way the burst durations are interpreted: in the case these events
belong to ultra-long GRBs~\cite{tikhomirova} the probability for
the mentioned set of events to be recurrent bursts is vanish.

Let us briefly discuss other effects that can arise when a WH
transmits a GRB. If the WH connects parts of our universe it is
possible to observe the same event twice from two distinct
directions. Indeed, one direction is source -- observer while the
other is WH -- observer. Profiles of these events have to be
identical within a time scale factor defined by the mass ratio of
the WH throats, with the first event being arbitrarily retarded
with respect to the second one. Such successive events coming from
the same point in space were proposed to be ordinary gravitational
lensing candidates (\cite{Paczynski, Mao}).

Indeed, extended literature is devoted to search for GRB with
similar light curves (see numerous papers dealing with the search
for classical recurrent GRBs and gamma-ray bursts with identical
light curves \cite{hurley_GRBrepeat, Wang1993,
Wang1995,Bennett1995,Band1996,Cline1996,Gorosabel1998,
Singhl1998,Wambsganss1993,McBreen2004}).  However there is no any
evidence that two or more bursts with similar light curve came
from the same source.

\section{Conclusion}
\label{s6}

In the paper we have considered a possibility in principle to
observe GRBs through a wormhole throat. We have obtained
theoretical power spectra (${\log N-\log s}$) and considered
observational properties of GRBs transmitted by a WH. Most
probably, the transient sources in gamma-ray domain might be
candidates in observable wormholes. However at present time there
is no candidates related to GRB transmission by wormhole.

One can guess what precisely are the events that could possibly be
identified as WH candidates. These are sources of repeating bursts
from the same source that are harder and longer than SGRs. These
could also be GRBs with similar light curves however from
different localization areas.

In order to observe a gamma-ray burst transmitted through a WH for
sure, we need the wormhole to be ${1\div 5}$ pc away from the
Earth. Alternatively, the gamma-ray burst source could be located
near the WH  throat in the other universe or in a distant part of
our universe. Then observations of WHs are possible at
cosmological distances. The candidates could be suspected, in the
former case, in GRBs with undetected host galaxies and, in the
latter case, in bright gamma-ray bursts with no optical afterglow
(so called dark GRB).

\section*{Acknowledgements}
\label{s-end}

We would like to thank N.S. Kardashev and O.V. Khoruzhii for
discussing this paper and providing valuable comments.

\bigskip
The work was supported by the Origin and Evolution of Stars and
Galaxies 2008 Program of Russian Academy of Sciences. AS
acknowledges the support of the Russian Foundation for Basic
Research (project codes: {07-02-01128-a}, {08-02-00090-a}) and of
scientific schools {NSh-626.2008.2}, {NSh-2469.2008.2}.

\end{document}